\documentclass[12pt]{article}

\usepackage{siunitx}
\usepackage{mathrsfs}  
\usepackage{units}
\usepackage{bbm}
\usepackage{multicol}

\usepackage{graphicx}
\usepackage{tabularx}
\usepackage{a4}
\usepackage{url}
\usepackage{longtable, lscape}
\usepackage[usenames, dvipsnames]{color}
 \usepackage{pdflscape}

\usepackage{multirow, makecell, bigstrut, booktabs, float}
\usepackage{array}

\usepackage{setspace}
\usepackage[fleqn]{amsmath}
\usepackage{amsfonts}
\usepackage{amssymb}
\usepackage{subfigure}
\usepackage{tcolorbox}
\usepackage{mathrsfs}  
\usepackage{bbm}

\usepackage{adjustbox}
\usepackage{nicefrac}
\usepackage{pgfplots}
\usepackage{authblk}
\begin{document}
\title{Causal inference with multi-state models -- estimands and estimators of the population-attributable fraction}

\author[1,2]{Maja von Cube\thanks{cube@imbi.uni-freiburg.de, Ernst-Zermelo Strasse 1, 79104 Freiburg}}

\author[1,2]{Martin Schumacher}
\author[1,2]{Martin Wolkewitz}

\affil[1]{Institute of Medical Biometry and Statistics, Faculty of Medicine and Medical Center - University of Freiburg, Freiburg, Germany}

\affil[2]{Freiburg Center for Data Analysis and Modelling, University of Freiburg, Freiburg, Germany}

\renewcommand\Authands{ and }

\date{March 8, 2019}

\maketitle


\begin{abstract}

The  population-attributable fraction (PAF) is a popular epidemiological measure for the burden of a harmful exposure within a population. It is often interpreted causally as proportion of preventable cases after an elimination of exposure. Originally, the PAF has been defined for cohort studies of fixed length with a baseline exposure or cross-sectional studies.

An extension of the definition to complex time-to-event data is not straightforward. We revise the proposed approaches in literature and provide a clear concept of the PAF for these data situations. The conceptualization is achieved by a proper differentiation between estimands and estimators as well as causal effect measures and measures of association.

\textbf{keywords:}Causal effect measure, Competing risks, Excess fraction, Measure of association, Nosocomial infection, Time-dependent exposure
\end{abstract}

\section{Introduction}

A relevant statistical quantity in epidemiology is the population-attributable fraction (PAF). Initially, the PAF has been defined for cross-sectional studies and cohort studies of fixed length with a baseline exposure. It summarizes the population-attributable burden by taking both the relative risk (RR) of exposure and the prevalence into account.
Provided adequate consideration of confounding the PAF is interpretable as a causal effect measure and quantifies the benefit of a preventive intervention against a harmful exposure.

However, in hospital epidemiology researchers are confronted with complex time-to-event data. To study, for example, the burden of hospital-acquired infections (HAIs) in terms of intensive care mortality, one has to account for the time-dependency of an acquisition of HAIs. All patients are naturally unexposed at intensive care unit (ICU) admission. Then, an infection may occur over the course of the ICU stay of the patients.

Additionally, discharge alive is a competing event to death in the ICU. Disregarding the competing risk by e.g. censoring discharged patients can lead to a strong overestimation of ICU mortality (\cite{schumacher2013hospital}).

The PAF being interpreted as proportion of preventable cases is commonly considered as causal effect measure (\cite{mansournia2018population}). Causal modelling of the effect of a harmful time-dependent exposure is challenging. For example, the different times of onset of the infection give rise to a high range of possible exposure paths. These are additionally complicated by the competing risks setting resulting in varying lengths of ICU stays and therefore varying lengths of follow-up.

Literature provides two different approaches to obtain the PAF for time-dependent exposures and competing outcomes. However, an extension of the PAF to a time-dependent setting is not straightforward. This article is devoted to a comprehensive revision of the literature on the PAF for complex time-to-event data arising in hospital epidemiology.

In Section \ref{sec:PAFs}, we review the definition of the PAF based on the extended illness-death model as proposed by \cite{schumacher2007attributable}. The definition is an extension of the time-independent estimand of the PAF provided by \cite{benichou2001review}. Statistical nomenclature allows for a differentiation between measures of association and causal effect measures. In this sense, \cite{benichou2001review} defined the PAF as measure of association that is based on observable quantities.

Section \ref{sec:PAF_Bmsm} is meant to view the PAF for time-varying exposures from a causal perspective. Causal effect measures can be defined by counterfactual outcomes. Such an approach has been proposed by \cite{bekaert2010adjusting}. Motivated by their work, we define an estimand that is a natural extension of the time-independent definition of the PAF as proposed by \cite{sjolander2011estimation}. In contrast to \cite{benichou2001review}, \cite{sjolander2011estimation} provides an estimand of the PAF that is based on counterfactual outcomes. Nonetheless, in the absence of confounding the time-independent definitions of the PAF by \cite{sjolander2011estimation} and \cite{benichou2001review} are equivalent (\cite{hernan2004definition}).
For identification of our proposed estimand, we use the same multi-state model as \cite{schumacher2007attributable}. The multi-state model approach allows for a direct comparison of the two time-dependent estimands of the PAF.

Such a comparison is performed in Section \ref{sec:comp}. We explain the difference between the two approaches and provide a framework which allows for a proper interpretation of both of the estimands. 

Then, based on this conceptualization of the two estimands, we revise the estimation procedures of the PAF as outlined by \cite{bekaert2010adjusting}. This is done in Section \ref{sec:InvB} by showing mathematically the proper use of their estimators. Finally, the approaches are applied to a sample of ICU patients to estimate the burden of hospital-acquired pneumonia (HAP). The data example illustrates the practical behaviour of the estimands and explains the interpretation based on a real data example. The article ends with a discussion.

\section{The PAF for time-dependent exposures and competing risks - defined with observable variables}\label{sec:PAFs}

\cite{benichou2001review} defined the PAF for cohort studies of fixed length with a baseline exposure and cross-sectional studies as 
\begin{equation}
PAF_o=\frac{P(D=1)-P(D=1|E=0)}{P(D=1)},
\label{eq:PAF_beni}
\end{equation}
where $E$ indicates whether the patient was exposed (for example infected) and $D$ indicates whether the patient experienced the outcome of interest (for example death in the ICU). 
This definition is based on observable random variables of exposure and outcome. Defined as measure of association, $PAF_o$ is interpretable as the proportion of attributable cases observed in association with exposure. In the absence of confounding it can be also interpreted as proportion of preventable cases if exposure was eliminated for all patients (\cite{walter1976estimation}). The index 'o' in the definition denotes that $PAF_{o}$ is defined with observable quantities.

In the following, we refer to cross-sectional studies and cohort studies of fixed length with a baseline exposure as 'time-fixed' data settings.

\subsection{The estimand $PAF_o(t)$}

\cite{schumacher2007attributable} extended the definition of the PAF proposed by \cite{benichou2001review} to account for the time-dependency of exposure and outcome. As \cite{schumacher2007attributable}, we use the example of HAIs to define the estimand. Nevertheless, it is applicable to any data situation with binary time-dependent exposure.
The estimand $PAF_o(t)$ is formally given by
\begin{equation}
PAF_o(t)=\frac{P(D(t)=1)-P(D(t)=1|E(t)=0)}{P(D(t)=1)},
\label{eqPAF_o}
\end{equation}
where $D(t)$ indicates if the patient died in the ICU by time $t$ and $E(t)$ if the patient acquired the exposure (i.e. an HAI) by time $t$. Both $E(t)$ and $D(t)$ are observable random variables. The variable $D(t)$ is equal to one if the patient was observed to die in the ICU by time $t$. Otherwise, if the patient was still alive at time $t$ (either discharged alive by $t$ or still in the ICU), then $D(t)=0$. Thus, $P(D(t)=1)$ is the overall ICU mortality risk in the patient population.

The death risk among unexposed, $P(D(t)=1|E(t)=0)$, is defined as a specific version of the conditional probability function (CPF) of  \cite{pepe1993kaplan}. At each time point $t$ the unexposed patients are defined to be those patients that did not acquire an infection until $t$. Therefore, they are an observable time-varying \textit{subpopulation} of the reference population. The CPF has been initially proposed as a summary measure for the probability of a certain outcome in the presence of competing risks. Considering the data example of HAIs, it is interpretable as the probability of death in the ICU by time $t$ given the patient remained uninfected until $t$. As discussed by \cite{pepe1993kaplan} and \cite{andersen2012interpretability}, the CPF is neither a predictive nor a causal quantity. Being based on observable quantities, $PAF_o(t)$ is defined as a measure of association. $PAF_o(t)$ relates the proportion of patients that acquired an HAI by time $t$ with that of the patients who remained HAI-free until time $t$.

\begin{figure}[hbtp!]
\centering
\includegraphics[width=\textwidth]{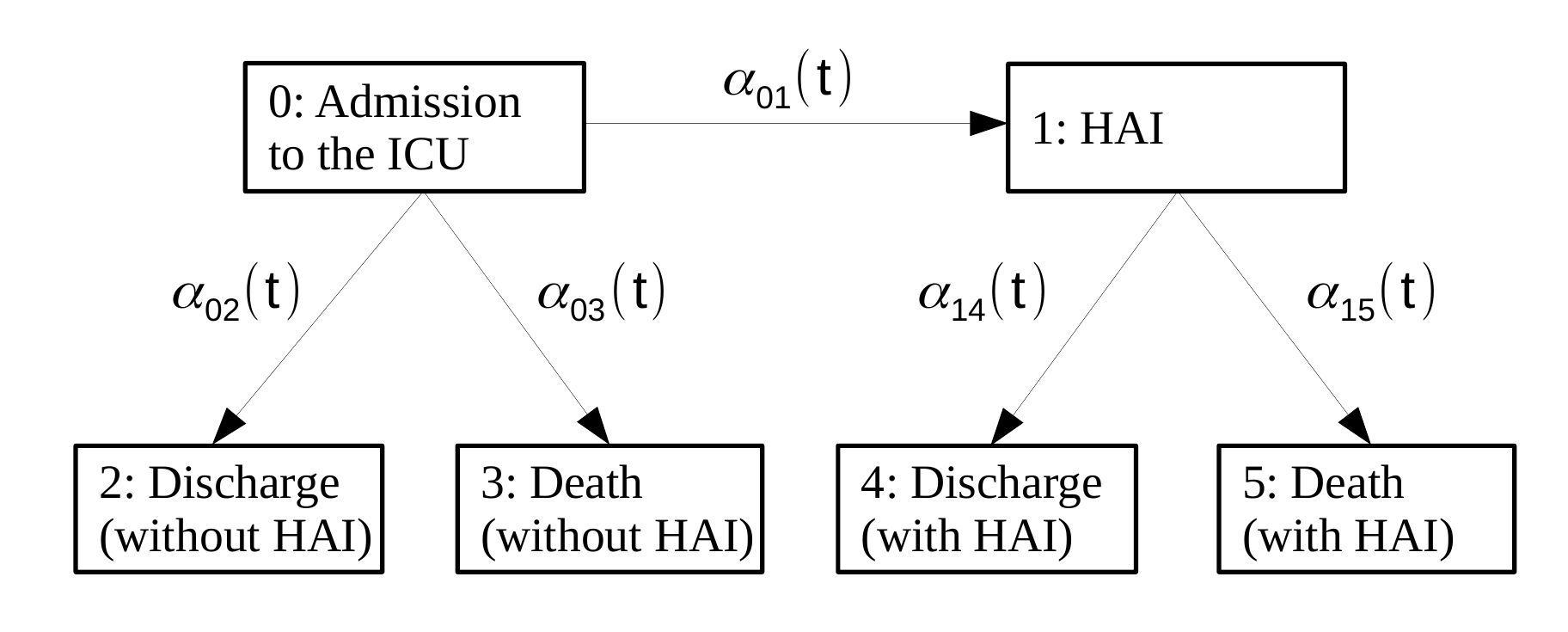}
 \caption{Extended illness-death model with hazard rates $\alpha_{01}(t)$, $\alpha_{02}(t)$, $\alpha_{03}(t)$, $\alpha_{14}(t)$ and $\alpha_{15}(t)$.}
  \label{fig:extendedIllnessDeath}
\end{figure}

As we regard only infections acquired in the hospital or ICU, all patients are free of HAI at the beginning of the study. This implies that $E(0)=0$ for all patients. Nonetheless, a relaxation of this property is straightforward and has no influence on the conclusions we are drawing.

\subsubsection{Estimation of $PAF_o(t)$}

The estimand $PAF_o(t)$ can be identified with the transition probabilities of the extended illness-death model (\cite{schumacher2007attributable}) shown in Figure \ref{fig:extendedIllnessDeath}. This model has the initial state 'Admission to the ICU' and accounts for HAI-acquisition as an intermediate state. Discharge alive from the ICU is a competing risk to death in the ICU. These two absorbing states are modelled both before and after HAI-acquisition. 
The risk of death at time $t$ among unexposed patients at $t$ is defined by
\begin{align}
P(D(t)=1|E(t)=0)&=\frac{P_{03}(0,t)}{P_{00}(0,t)+P_{02}(0,t)+P_{03}(0,t)}\label{condP03}\\
&=\frac{P_{03}(0,t)}{1-(P_{01}(0,t)+P_{04}(0,t)+P_{05}(0,t))},\nonumber
\end{align}
where $P_{0l}(0,t)$ ($l=0,2,3$) denote the transition probabilities of the extended illness-death model. This quantity is referred to as CPF (\cite{pepe1993kaplan}).
More explicitly, the transition probabilities are defined as
\begin{align}
P_{00}(0,t)&=\exp\Bigl(-\int_{0}^{t}\alpha_{01}(u)+\alpha_{02}(u)+\alpha_{03}(u)~du\Bigr)\nonumber \\ 
&=\exp\Bigl(-\int_{0}^{t}\alpha_0(u)~du\Bigr),\nonumber \\
P_{02}(0,t)&=\int_{0}^{t}P_{00}(0,u)\alpha_{02}(u)~du,\label{eq:transProb}\\
P_{03}(0,t)&=\int_{0}^{t}P_{00}(0,u)\alpha_{03}(u)~du, \nonumber\\
\nonumber
\end{align}
where $\alpha_{kl}(t)$, $k=0,1$, $l=1,2,3,4,5$, are the cause-specific hazard rates of the extended illness-death model and $\alpha_0(t)=\alpha_{01}(t)+\alpha_{02}(t)+\alpha_{03}(t)$. 
In contrast to the marginal probability $P_{03}(0,t)$, the conditional probability $P(D(t)=1|E(t)=0)$ \eqref{condP03} conditions on the additional information that patients remained infection-free until time $t$.

The overall ICU death risk is identified by
\begin{equation}
P(D(t)=1)=P_{03}(0,t)+P_{05}(0,t),
\label{P_Dt}
\end{equation}
with $P_{05}(0,t)$ being explicitly defined as
\begin{align}
P_{05}(0,t)&=\int_{0}^{t}P_{00}(0,u)\alpha_{01}(u)P_{15}(u,t)~du,\label{eq:transProb05}\\
P_{15}(0,t)&=\int_{0}^{t}P_{11}(0,u)\alpha_{15}(u)~du,\\
P_{11}(0,t)&=\exp\Bigl(-\int_{0}^{t}\alpha_{14}(u)+\alpha_{15}(u)~du\Bigr).\nonumber\\
\nonumber
\end{align}

Therefore, $P(D(t)=1)$ depends on all five hazard rates of the extended illness-death model. The death risk among unexposed, $P(D(t)=1|E(t)=0)$, depends on the death and discharge hazard without an HAI as well as on the infection hazard.

Estimation of $PAF_o(t)$ is based on the Aalen-Johansen estimators of the transition probabilities of the extended illness-death model. For details, we refer to \cite{beyersmann2011competing} and \cite{schumacher2007attributable}. 
When using the Aalen-Johansen estimators for $P_{0l}(0,t)$ ($l=0,2,3,5$), the Markov assumption is assumed to be plausible. For the extended illness-death model, the Markov assumption means that the risk of death and discharge given an HAI depends only on the total length of stay in the hospital and not on the length of stay in the hospital until acquisition of an HAI. The plausibility of this assumption can be tested by including time of acquisition of an HAI as covariate in a Cox proportional hazards model for the transitions from State 1 to States 4 and 5 (\cite{keiding1990random}).

Nevertheless, $PAF_o(t)$ can be estimated without having to make this assumption. The overall mortality risk $P(D(t)=1)$ can be equivalently identified by a competing risks model in which States 2 and 4 and States 3 and 5 are combined. The resulting model does not differentiate between infected and uninfected patients but validly models overall ICU mortality. In competing risks models, the Markov assumption is naturally avoided (\cite{beyersmann2011competing}).

The death risk among unexposed, $P(D(t)=1|E(t)=0)$, can be estimated by a three-states competing risks model which results from Figure \ref{fig:extendedIllnessDeath} by dropping States 4 and 5. With this model, $P(D(t)=1|E(t)=0)$ can be identified since $P_{03}(0,t)$ is independent of the events past exposure acquisition and the sum $P_{01}(0,t)+P_{04}(0,t)+P_{05}(0,t)$ is equal to the cumulative incidence function (CIF) of acquiring an HAI. Both of these probabilities can be identified and estimated with a competing risks model. For more details, we also refer to the appendix.

An adjusted $PAF_o(t)$ can be obtained either by stratification based on baseline covariates or by using a semi-parametric model as proposed by \cite{coeurjolly2012attributable}. For the estimation of 95\% confidence intervals (CIs) we suggest to use a bootstrap method (\cite{efron1986bootstrap}).

\section{The PAF for time-dependent exposures and competing risks - defined with counterfactual variables}\label{sec:PAF_Bmsm}

The PAF as a causal effect measure for time-fixed data settings has been defined by \cite{sjolander2011estimation} as
\begin{equation}
PAF_c=\frac{P(D=1)-P(D_0=1)}{P(D=1)},
\label{eq:PAF_sjo}
\end{equation} 
where $P(D=1)$ is the observable risk of the outcome and $P(D_0=1)$ the hypothetical risk had the patients remained -- possibly contrary to fact -- unexposed. $PAF_c$ is interpretable as the proportion of preventable cases if exposure was eliminated for all patients. In the absence of confounding $PAF_c$ and $PAF_o$ are equivalent (\cite{hernan2004definition}). The index 'c' in the definition of $PAF_{c}$ refers to the counterfactual nature of the estimand.

\subsection{The estimand $PAF_c(t)$}

In Section \ref{sec:PAFs}, the PAF has been identified using a multi-state model that accounted for infection as competing event to death without infection. This approach was based on observable quantities which are commonly used to make predictions.

Yet, the PAF is most often interpreted as a causal effect measure (\cite{mansournia2018population}). Considering our data example it aims at answering the question of how many death cases occurring in the ICU would be spared if an intervention prevented HAIs.

\cite{bekaert2010adjusting} proposed to use causal inference to estimate the PAF. We formalize their approach and define the estimand with the desired interpretation by
\begin{equation}
PAF_c(t)=\frac{P(D(t)=1)-P(D_0(t)=1)}{P(D(t)=1)},
\label{eqPAF_B}
\end{equation}
where $D(t)$ is defined in the same way as in Section \ref{sec:PAFs}, namely as the observable random variable of death by time $t$. The variable $D_0(t)$ is the counterfactual outcome at $t$ had the patient been (possibly contrary to fact) unexposed until time $t$. Thus, $P(D_0(t)=1)$ is the hypothetical death risk at $t$ had all patients remained unexposed until time $t$ or the end of their ICU stay, whichever comes first. This definition of the PAF compares the same patient population under two distinct conditions, namely the population as it is actually observed compared to the same population in a world without infections. Hence, $PAF_c(t)$ is interpretable as proportion of preventable cases if exposure was extinct.

\subsubsection{Estimation of $PAF_c(t)$ using a multi-state model}\label{sec:PAFBb}

We use the extended illness-death model in Figure \ref{fig:extendedIllnessDeath} to identify $PAF_c(t)$. The overall mortality risk, $P(D(t)=1)$, is the same statistical quantity as in the definition of $PAF_o(t)$. Therefore, it can be identified in the same way as explained in Section \ref{sec:PAFs}.

\begin{figure}[hbt!]
\centering
\includegraphics[width=13cm]{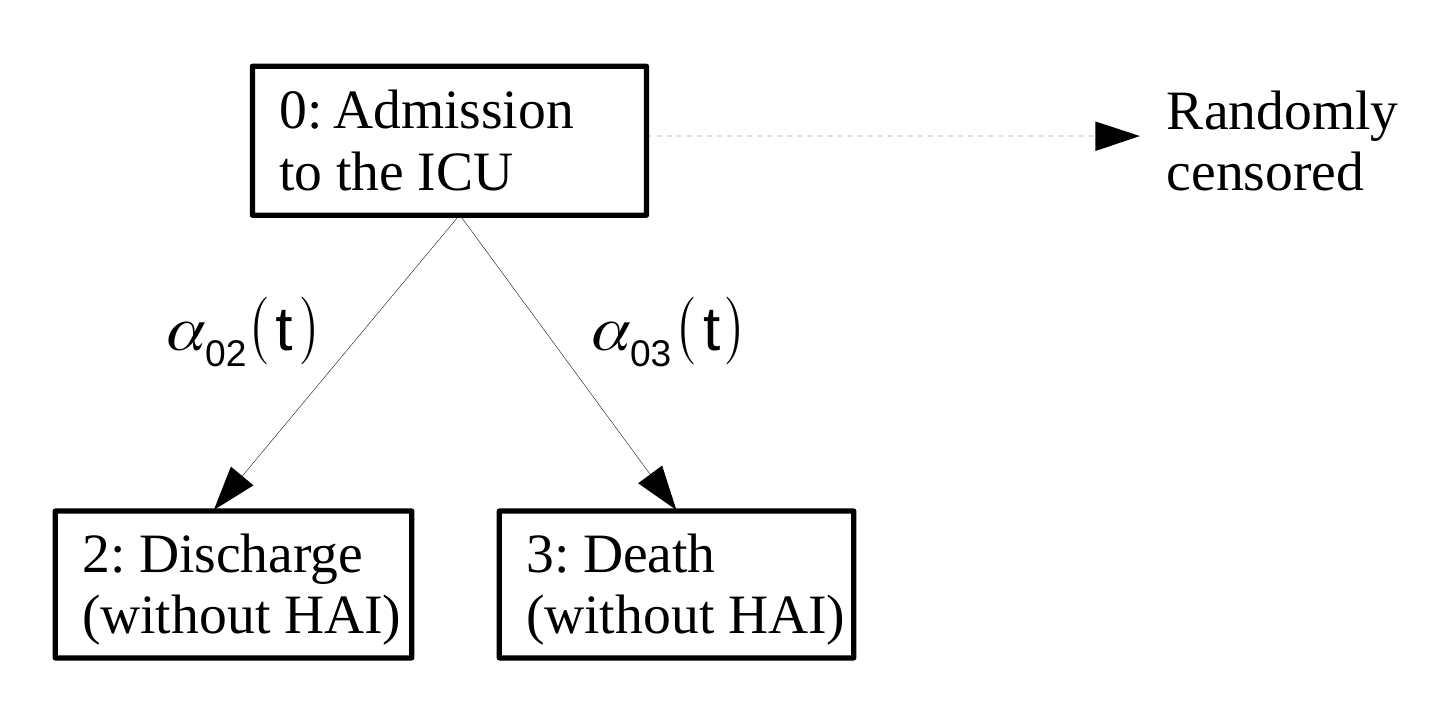}
\caption{The competing risks model with cause-specific hazard rates $\alpha_{02}(t)$, $\alpha_{03}(t)$. In contrast to Figure 5 in the appendix, patients that acquire an HAI are randomly censored with censoring rate $\alpha_{01}(t)$. The event 'Acquisition of HAI' is not being considered.}
\label{fig:CRcens}
\end{figure}

The hypothetical death risk had all patients remained unexposed until time $t$ or the end of their ICU stay, $P(D_0(t)=1)$, can be identified with the extended illness-death model by setting the infection hazard to zero. The resulting model is a competing risks model with the two states death and discharge without an HAI (see Figure \ref{fig:CRcens}). 
Then, $P(D_0(t)=1)$ is identified by the CIF of death without an HAI. We denote the transition probabilities of this model with $P_{0l_0}(0,t)$, $l=0,2,3$.
They are explicitly given by
\begin{align}
P_{00_0}(0,t)&=\exp\Bigl(-\int_{0}^{t}\alpha_{02}(u)+\alpha_{03}(u)~du\Bigr),\nonumber \\ 
P_{02_0}(0,t)&=\int_{0}^{t}P_{00_0}(0,u)\alpha_{02}(u)~du,\label{eq:transProb0}\\
P_{03_0}(0,t)&=\int_{0}^{t}P_{00_0}(0,u)\alpha_{03}(u)~du.\nonumber\\
\nonumber
\end{align}
The estimands $P_{0l_0}(0,t)$, $l=0,2,3$, are independent of the infection hazard.

An estimator of $P_{03_0}(0,t)$ is for example the Aalen-Johansen estimator in the competing risks model (Figure \ref{fig:CRcens}). In contrast to the transition probability $P_{03}(0,t)$ obtained from the extended illness-death model (Figure \ref{fig:extendedIllnessDeath}), patients acquiring an infection are treated as censored observations. We denote the Aalen-Johansen estimator of $P(D_0(t)=1)$ by $\hat{P}_{03_0}(0,t)$. 

The proposed estimation procedure for $PAF_c(t)$ makes use of an inference technique for competing risks which results by assuming equivalence of a competing risk and independent censoring. More precisely, independent censoring means that the cause-specific hazard rates remain undisturbed by censoring (\cite{andersen1993statistical}). 
Generally, it is strongly disadvised to handle competing risks as independently censored events (e.g. \cite{beyersmann2011competing, andersen1993statistical, kalbfleisch1980statistical}). Even if the assumption of independence is justified, the estimands that result from this estimation procedure are considered to have a questionable interpretation (see e.g. \cite{beyersmann2011competing, andersen1993statistical, kalbfleisch1980statistical}). Nevertheless, they appear to be useful for defining and estimating the PAF: The population parameters, $P_{{0l}_0}(s,t)$, $l=0,2,3$, are meant to be derived for the same population as the estimands $P_{0l}(s,t)$, $l=0,2,3$. However, they have a different interpretation. While $P_{03}(s,t)$ is the risk of death by time $t$ without an HAI, $P_{{03}_0}(s,t)$ is the risk of death by time $t$ if HAIs were prevented. Thus, $P_{03}(s,t)$ is a probability of factual events, while $P_{{03}_0}(s,t)$ is a probability of hypothetical ones. Censoring a competing risk allows for an investigation of hypothetical scenarios (\cite{keiding2001multi}).

In our setting the independent censoring assumption translates into assuming that patients who acquired an HAI are under the same risk of death and discharge as patients who are still infection-free in the ICU. Hence, the proposed estimator is unbiased if patients that remain infection-free are representative of all patients. This assumption is called exchangeability assumption. In an observational study, it is highly unlikely that exposed and unexposed patients are exchangeable.  Therefore, the estimator must be adjusted for confounding. Due to the time-dependency of exposure confounding is generally time-varying as well and feedback between the confounders and exposure is to be expected (\cite{mansournia2017handling}). Adjustment for such confounding can be done by using inverse-probability weights (\cite{mansournia2016inverse}). 
\cite{bekaert2010adjusting} provide a method which allows for adjustment for confounding of $PAF_c(t)$ in a time-discrete approach. This methodology is discussed in more detail in Section \ref{sec:InvB}.
Moreover, \cite{bekaert2010adjusting} shortly discuss the assumptions of consistency and positivity which are fundamental to identify causally interpreted effect measures from observational data (see also the appendix). A more detailed definition of these assumptions is provided by \cite{hernan2018estimate}. For estimation of the 95\%-CIs, we propose to use a bootstrap approach (\cite{efron1986bootstrap}).

\section{Conceptualization of $PAF_o(t)$ and $PAF_c(t)$}\label{sec:comp}

Both $PAF_o(t)$ and $PAF_c(t)$ extend the time-independent definition of the PAF. $PAF_o(t)$ is defined by observable quantities and is an extension of the definition provided by \cite{benichou2001review}. In contrast, $PAF_c(t)$ is defined with counterfactuals and extends the definition by \cite{sjolander2011estimation}. 

In the time-fixed data setting, we can achieve -- under certain assumptions -- equivalence of $PAF_o$ and $PAF_c$ (\cite{hernan2004definition}). In fact, the causally defined effect measure $PAF_c$ is identifiable only if $\widehat{PAF}_o$ has a causal interpretation. Then, $\widehat{PAF}_o$ estimates $PAF_c$. For a detailed explanation under which conditions $PAF_c$ is identifiable, we refer to \cite{sjolander2011estimation} .

Comparing $PAF_o(t)$ and $PAF_c(t)$, we find that this basic relation of causal effect measures and measures defined as associational quantities no longer holds in this way if exposure depends on time. While the factual overall mortality risk in the population, $P(D(t)=1)$, is the same in both definitions, the two estimands, $PAF_o(t)$ and $PAF_c(t)$, differ in the definition of the death risk among unexposed patients. First, we note that generally
\begin{equation*}
P(D_0(t)=1)\neq P(D(t)=1|E(t)=0).
\end{equation*}

Even in the absence of confounding, the death risk in the subpopulation of patients that were observed to be unexposed until time $t$, $P(D(t)=1|E(t)=0)$, is not the same as that in the hypothetical population in which all patients would have remained unexposed until $t$, $P(D_0(t)=1)$. This is in contrast to the time-fixed situation, where these two quantities are indeed the same. 

\cite{shmueli2010explain} explain the difference between explanatory modelling which aims at understanding causal relations and descriptive modelling which aims at summarizing the data structure and illustrating basic associations. Based on this conception of statistical modelling, we now provide a clear concept of the two estimands and their correct interpretation. 

When taking the \textbf{causal point of view}, the goal is explanatory modelling of the PAF. The PAF interpreted as a causal effect measure (i.e. proportion of preventable cases, $PAF_c(t)$) aims to compare the factual patient population with a hypothetical one had all patients remained unexposed. However, the subpopulation of patients observed to be unexposed at time $t$ (patients with $(E(t)=0)$) is not a sample of this hypothetical population. This is even the case in the absence of confounding.

More precisely, the \textit{estimand} $P(D(t)=1|E(t)=0)$ is structurally biased from a causal perspective. The bias arises by conditioning on $E(t)=0$. By assessing the infection state specifically at $t$ while allowing for death within $(0,t]$, we condition upon having reached an absorbing state thereby creating selection bias (\cite{andersen2012interpretability}). The condition $E(t)=0$ translates into the condition of being in either State 0, 2 or 3, where States 2 and 3 are absorbing.

In other words, the hypothetical intervention that could prevent exposure is ill-defined. As a consequence, the estimand $PAF_o(t)$ is not interpretable as a causal effect measure. Hence, in contrast to $PAF_c(t)$, $PAF_o(t)$ is not interpretable as proportion of preventable cases if a hypothetical intervention eliminated exposure, even in the absence of confounding.


Thus, on the one hand, $PAF_o(t)$ extends the time-fixed definition of \cite{benichou2001review} as a basic relation of the proportion of cases. On the other hand, it is not an extension of the PAF as a causal effect measure. 

When taking the \textbf{observational point of view} the goal is descriptive modelling of the PAF. $PAF_c(t)$ is not a summary measure for the factual accumulation of attributable death cases associated with exposure. For example, HAIs occur over the course of time and are by definition not present at baseline. As a consequence, death cases accumulate first among unexposed and with a time-delay among exposed patients. While the time-delay is not a causal consequence of HAIs, it remains a natural characteristic that comes along with HAIs. For a proper understanding of the timing of exposure and outcome, the time of infection should be accounted for. $PAF_c(t)$ depends only indirectly on the time of infection via $P(D(t)=1)$. However, the intensity of HAI acquisition is not captured.

\section{Conceptualization of estimators for $PAF_o(t)$ and $PAF_c(t)$}\label{sec:InvB}

The failure to properly distinguish these two effect measures has caused misconceptions in literature. In the following, we clarify these misconceptions by proving that the two estimation procedures proposed by \cite{bekaert2010adjusting} lead to the estimation of two different estimands, namely $PAF_o(t)$ and $PAF_c(t)$. 

First, we introduce the estimation procedure proposed by \cite{bekaert2010adjusting}. They provide two estimators of the PAF based on estimated CIFs. As discussed in Section \ref{sec:comp}, the difference between $PAF_o(t)$ and $PAF_c(t)$ is via the definition of the death risk among unexposed. The observable overall mortality risk is equivalent in all approaches.

\cite{bekaert2010adjusting} denote the (estimated) death risk among unexposed by $\hat{F}_{\overline{0}1}(t)$ and define it as the "estimated counterfactual CIF of ICU mortality at time $t$ under the no-infection path" (\cite{bekaert2010adjusting}). The no-infection path implies that the patient remained -- possibly contrary to fact -- infection-free during the complete follow-up time $(0,\tau]$. The no-infection path is equivalent to assuming that the patient remained unexposed until the end of their ICU stay. For a detailed explanation of the hypothetical infection paths we refer to \cite{bekaert2010adjusting}. To properly differentiate between estimands and estimators, we denote the counterfactual CIF of ICU mortality at time $t$ under the no-infection path by $F_{\overline{0}1}(t)$.

\cite{bekaert2010adjusting} refer to the unadjusted version of their estimated PAF as naive, since it is not adjusted for confounding. This estimator is given by 

\begin{align*}
\widehat{F}_{\overline{0}1}^{naive}(t)&:=\frac{\sum_{s=1}^td_{\overline{0}1s}}{n_{\overline{0}t}}\\
&=\frac{\sum_{i=1}^n \mathbbm{1}(\epsilon_i(t)=1)\prod_{s=1}^t(\mathbbm{1}(A_i(s)=0)\mathbbm{1}(\epsilon_i(s-1)=0)+\mathbbm{1}(\epsilon_i(s-1)\neq0))}{\sum_{i=1}^n \prod_{s=1}^t(\mathbbm{1}(A_i(s)=0)\mathbbm{1}(\epsilon_i(s-1)=0)+\mathbbm{1}(\epsilon_i(s-1)\neq0))},
\end{align*}
where $n$ is the total number of sampled patients in the study, $\epsilon_i(t)$ is the observable outcome of patient $i$ at time $t$ with $\epsilon_i(t)=1$ if the patient $i$ died in the ICU by time $t$, $\epsilon_i(t)=2$ if the patient was discharged alive by $t$ and $\epsilon_i(t)=0$ if the patient was still in the ICU. The quantity $A_i(t)$ is an indicator of the observed infection state of patient $i$ at time $t$. 
We denote the estimator of the PAF resulting from using $\widehat{F}_{\overline{0}1}^{naive}(t)$ as "estimated counterfactual CIF of ICU mortality at time $t$ under the no-infection path" by $\widehat{PAF}_B^{naive}(t)$. 

Bekaert et al. propose to adjust for confounding by adjusting $\widehat{F}_{\overline{0}1}^{naive}(t)$ using inverse-probability of exposure weights (IPWs). This estimator is formally given by

\begin{equation*}
\hat{F}^{IPW}_{\overline{0}1}(t)=\frac{\sum_{i=1}^n \mathbbm{1}(\epsilon_i(t)= 1) W_{i\overline{0}t}}{\sum_{i=1}^nW_{i\overline{0}t}},
\end{equation*}
with
\begin{equation*}
W_{i\overline{0}t}=\prod_{s=1}^t\frac{(\mathbbm{1}(A_i(s)=0)\mathbbm{1}(\epsilon_i(s-1)=0)+\mathbbm{1}(\epsilon_i(s-1)\neq0)}{P(A_i(s)=0|(\epsilon_i(s-1), A_i(s-1))}.
\end{equation*}

We denote the estimator of the PAF that results from using $\hat{F}^{IPW}_{\overline{0}1}(t)$ as the "estimated counterfactual CIF of ICU mortality at time $t$ under the no-infection path" by $\widehat{PAF}^{IPW}_B(t)$.

\subsection{Properties of $\widehat{PAF}_B^{naive}(t)$ and $\widehat{PAF}^{IPW}_B(t)$}\label{sec:NAIVEvsIPW}

To investigate the estimation procedure of the PAF proposed by \cite{bekaert2010adjusting} more closely, we first consider $\widehat{F}_{\overline{0}1}^{naive}(t)=\frac{\sum_{s=1}^td_{\overline{0}1s}}{n_{\overline{0}t}}$.

Without censoring, the number of patients that are observed to be unexposed until time $t$ are those that were discharged alive or died without an HAI within $(0,t]$ and those that are still at risk and uninfected at $t$. For each of these patients -- an individual patient is denoted by $i$ -- we have 
\begin{align*}
\prod_{s=1}^t(\mathbbm{1}(A_i(s)=0)\mathbbm{1}(\epsilon_i(s-1)=0)+\mathbbm{1}(\epsilon_i(s-1)\neq0))=1\\
\end{align*}
for $t\leq \tau$. Otherwise, if the patient acquired an infection any time before or at $t$, the product for this patient is equal to zero. Thus,
\begin{align*}
n_{\overline{0}t}&=\sum_{i=1}^n \prod_{s=1}^t(\mathbbm{1}(A_i(s)=0)\mathbbm{1}(\epsilon_i(s-1)=0)+\mathbbm{1}(\epsilon_i(s-1)\neq0))\\
&=\text{\# of patients observed to be unexposed until $t$}
\end{align*}

Equivalently, we have for the numerator that 
\begin{align*}
\sum_{s=1}^td_{\overline{0}1s}&=\sum_{i=1}^n \mathbbm{1}(\epsilon_i(t)=1)\prod_{s=1}^t(\mathbbm{1}(A_i(s)=0)\mathbbm{1}(\epsilon_i(s-1)=0)+\mathbbm{1}(\epsilon_i(s-1)\neq0))\\
&=\text{\# of patients observed to have died without HAI by time $t$}.
\end{align*}

Thus, $\widehat{F}_{\overline{0}1}^{naive}(t)$ is the proportion of patients that died without exposure by time $t$ among those that remained unexposed until time $t$. 

Without independent censoring, the estimated death risk among unexposed of $PAF_o(t)$, $\widehat{P}(D(t)=1|E(t)=0)$, is also simply the proportion of patients that died without exposure by time $t$ among those that remained unexposed until $t$ (\cite{von2017basic}). 

Therefore, 
\begin{equation*}
\widehat{P}(D(t)=1|E(t)=0)=\widehat{F}_{\overline{0}1}^{naive}(t)
\end{equation*}
having as direct consequence that 
\begin{equation*}
\widehat{PAF}_o(t)=\widehat{PAF}^{naive}_B(t).
\end{equation*}

In the appendix, we provide a detailed proof of $\widehat{PAF}_B^{naive}(t)$ estimating $PAF_o(t)$. The proof is based on showing that $\widehat{F}_{\overline{0}1}^{naive}(t)=\widehat{P}(D(t)=1|E(t)=0)$.

Moreover, we proof in the appendix that the use of the weighted estimator $\hat{F}^{IPW}_{\overline{0}1}(t)$ results in an estimator of $PAF_c(t)$. In the absence of confounding, we have that 
\begin{equation*}
\hat{F}^{IPW}_{\overline{0}1}(t)=\hat{P}_{03_0}(t).
\end{equation*}
The proof is based on showing that the $\hat{F}^{IPW}_{\overline{0}1}(t)$ is equivalent to a Horvitz-Thompson like estimator of the CIF. However, instead of weights that are the inverse of the probability of being randomly censored, the weights are the inverse probability of exposure weights. Thus, patients that acquire an HAI are treated in the same way as if they were randomly censored. This shows that $\hat{F}^{IPW}_{\overline{0}1}(t)=\hat{P}_{03_0}(t)$ and therefore 

\begin{equation*}
\widehat{PAF}_c(t)=\widehat{PAF}^{IPW}_B(t).
\end{equation*}

Hence, even in the absence of confounding, we generally have
\begin{equation*}
\widehat{PAF}^{naive}_B(t)\neq\widehat{PAF}^{IPW}_B(t).
\end{equation*}
This implies especially that $\widehat{PAF}^{naive}_B(t)$ and $\widehat{PAF}^{IPW}_B(t)$ differ in interpretation. The two estimators estimate two distinct statistical quantities.
The unadjusted version, $\widehat{PAF}^{naive}_B(t)$, is an estimator of $PAF_o(t)$. The adjusted version, $\widehat{PAF}^{IPW}_B(t)$, is an estimator of $PAF_c(t)$ even in the absence of confounding. 
We highlight that the difference between $\widehat{PAF}^{naive}_B(t)$ and $\widehat{PAF}^{IPW}_B(t)$ is not solely due to confounding as explained by \cite{bekaert2010adjusting}.

In Table \ref{tab:estPAF_B}, we summarize the estimands and according estimators.

\begin{table}
\begin{center}
\begin{tabular}{l l l l}
\textbf{Estimand} & \textbf{Modelling } & \textbf{Interpretation} & \textbf{Estimator}\\
& \textbf{intention} &&\\
&&&\\
$PAF_o(t)=$ & Descriptive & Proportion of & $\widehat{PAF}_o(t)$, \\ 
$\frac{P(D(t)=1)-P(D(t)=1|E(t)=0)}{P(D(t)=1)}$&&  observable attributable & $\widehat{PAF}_B^{naive}(t)$ \\
&&cases accumulated &\\
&&up to time $t$&\\
&&&\\
$PAF_c(t)=$& Causal/explanatory &Proportion of  &  $\widehat{PAF}_c(t)$, \\
$\frac{P(D(t)=1)-P(D_0(t)=1)}{P(D(t)=1)}$ & &preventable cases & $\widehat{PAF}_B^{IPW}(t)$ \\
&&up to time $t$&\\
&&&\\
&&&\\
$P(D(t)=1|E(t)=0)$ & Descriptive & Proportion of cases until $t$ & $\frac{\hat{P}_{03}(t)}{\hat{P}_{00}(t)+\hat{P}_{02}(t)+\hat{P}_{03}(t)}$, \\
&& among unexposed at $t$ & $\widehat{F}_{\overline{0}{1}}^{naive}(t)$\\ 
&&&\\
&&&\\
$P(D_0(t)=1)$& Causal/explanatory  & Hypothetical proportion & $\hat{P}_{{03}_0}(t)$,$\widehat{F}_{\overline{0}{1}}^{IPW}(t)$\\
&&of cases until $t$ &\\ 
&& had all patients  &  \\
&&remained unexposed&\\
&&&\\
\end{tabular}
\caption{The various estimators of $PAF_o(t)$, $PAF_c(t)$, $P(D(t)=1|E(t)=0)$ and $P(D_0(t)=1)$.}
\label{tab:estPAF_B}
\end{center}
\end{table}

\section{Data example}

In this section, we study the PAF of hospital-acquired pneumonia (HAP). 
We estimate $PAF_c(t)$ to quantify the benefit of an intervention that could prevent HAP for all patients admitted to the ICU during their complete ICU stay. Moreover, we estimate $PAF_o(t)$ to understand the population-attributable burden of HAP. It provides an explanation for clinicians about how death cases accumulate over the course of time among uninfected patients in relation to the number of cases occurring among all patients. $PAF_o(t)$ summarizes what clinicians see on their daily routine.

The data is a sample of the SIR-3 (Spread of nosocomial Infection and Resistant pathogens) study (see \cite{wolkewitz2008risk}) and similar to the data example of \cite{schumacher2007attributable}.
The SIR-3 sample is publicly available in the R-package \textit{kmi} (\cite{kmi2017}). 
It consists of 1313 patients of whom 108 eventually acquired a HAP. Of the patients who remained free of HAP 126 died in the ICU and 1062 were discharged alive. Seventeen patients were censored before acquiring a HAP. Of the patients who were observed to be infected, 21 died in the ICU, 82 were discharged alive and 22 were lost to follow-up. The one patient who remained for more than 100 days in the ICU was administratively censored after 100 days. We first studied the effect of HAP on the death and discharge rates by considering HAP as a time-dependent covariate in a Cox proportional hazards model (Cox PH model). As typically found in the study of HAIs in ICUs in Europe, see e.g. \cite{beyersmann2011competing} and \cite{vonCube2018relative}, there was no direct effect of HAP on the death hazard (HR=0.99; 95\%-CI $[0.61; 1.60]$). However, the discharge hazard was significantly decreased for patients with HAP (HR=0.61; 95\%-CI $[0.48; 0.76]$). As discussed by e.g. \cite{vonCube2018relative}, this means that patients with HAP stay longer in the ICU and therefore longer at risk to die in the ICU. 

\begin{figure}[hbt!]
\centering
\includegraphics[width=\textwidth]{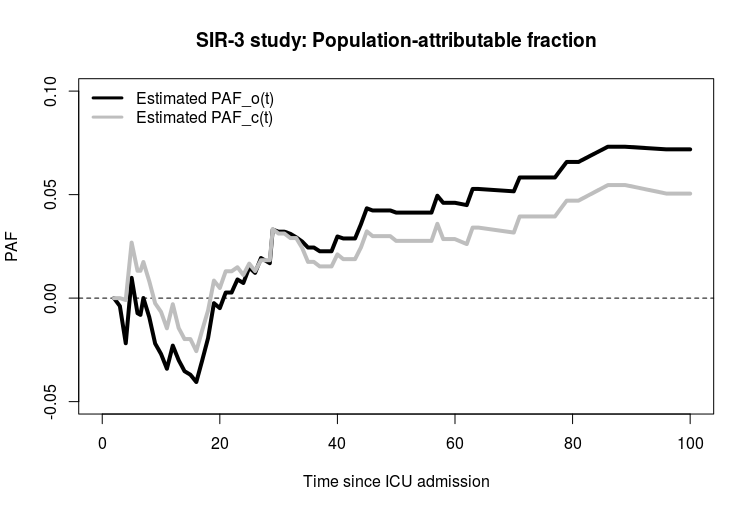}
\caption{$\widehat{PAF}_o(t)$ and $\widehat{PAF}_c(t)$ of HAP for a sample of 1313 patients of the SIR-3 study.}
\label{sir3paf}
\end{figure}

To understand the burden of HAP on a population level and the benefit of a preventive intervention, we estimate both $PAF_o(t)$ and $PAF_c(t)$ using the described methods. The result is shown in Figure \ref{sir3paf}. Adjustment for the available time-independent covariates age and gender did not alter the estimators in a mentionable way.

Both, $\widehat{PAF}_c(t)$ and $\widehat{PAF}_o(t)$ vary around zero within the first 20 days since ICU admission. Then, both continuously increase. 

After 100 days $\widehat{PAF}_o(t)$ is 0.073 explaining that 7.3\% or respectively 11 death cases out of all death cases that occurred by then were observed to be excess in association with HAP.

The proportion of preventable cases after 100 days, $\widehat{PAF}_c(100)$, is 0.055. This corresponds to 8 preventable cases if an intervention eliminated HAP during the complete ICU stay of the patients or up to 100 days, whichever occurred first. 

\begin{figure}[hbt!]
\centering
\includegraphics[width=\textwidth]{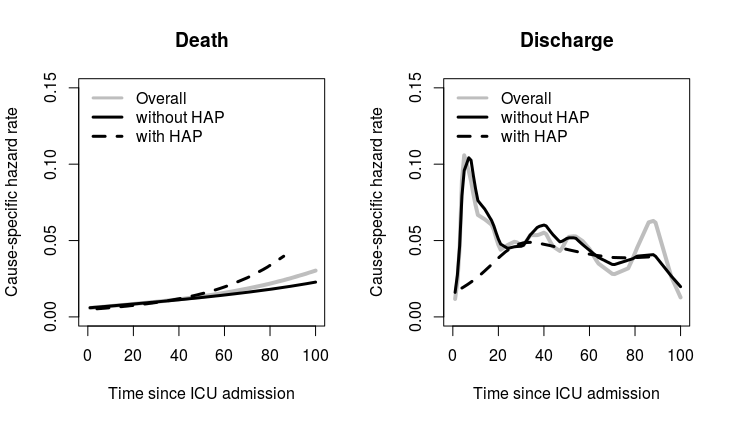}
\caption{Cause-specific hazard rates of the sample of 1313 patients of the SIR-3 study.}
\label{sir3hazards}
\end{figure}

In this data sample, $\widehat{PAF}_o(t)$ becomes greater than $\widehat{PAF}_c(t)$ for larger $t$. This is not generally the case. In a recently submitted manuscript (\cite{vonCube2018simulation}) we studied various simulated scenarios to investigate the behaviour of $PAF_o(t)$ and $PAF_c(t)$. 
A typical scenario in which $PAF_o(t)$ is greater than $PAF_c(t)$ is a setting, where the discharge hazard without HAP is large in the beginning and then decreases while at the same time, the death hazard without HAP increases from a low level. Such a scenario is comparable to the hazard rates of our data sample shown in Figure \ref{sir3hazards}. While $P(D_0(t)=1)$ depends only on the death and discharge hazards without HAP, $P(D(t)=1|E(t)=0)$ depends also on the infection hazard. The overall death risk $P(D(t)=1)$ depends on all the five hazard rates. 

$PAF_o(t)$ is typically below zero within the first days, as patients may already die without a HAP while it takes some time until the first patients have acquired a HAP to then die with it (\cite{von2017basic}). Both $\widehat{PAF}_o(t)$ and $\widehat{PAF}_c(t)$ vary around zero as the death hazard with HAP is slightly smaller than the death hazard without a HAP in the first 20 days. At the same time the discharge hazard without HAP is much larger but strongly decreasing between day 10 and 20, while the discharge hazard with HAP increases. As $\widehat{PAF}_o(t)$ and $\widehat{PAF}_c(t)$ are both cumulative measures, and the overall effect of HAP is harmful, both $PAF_B(t)$ and $PAF_o(t)$ eventually continuously increase. The behaviour of $\widehat{PAF}_o(t)$ and $\widehat{PAF}_c(t)$ demonstrates the sensitive dependence on the five hazard rates. We remark, that the difference between $\hat{P}(D(t)=1|E(t)=0)$ and $P(D_0(t)=1)$ is considerable small, due to the low infection hazard.

\section{Discussion}


In this article, we revised and advanced the literature on the PAF for data settings commonly arising in hospital epidemiology. By a mathematical exploration, we showed that literature provides two distinct definitions which we have denoted by $PAF_o(t)$ and $PAF_c(t)$.  Definition and estimation of $PAF_o(t)$ was a revision of the proposition by \cite{schumacher2007attributable}. Definition of $PAF_c(t)$ was motivated by the work of \cite{bekaert2010adjusting}. The estimation procedure, we proposed for $PAF_c(t)$ has been suggested in a technical report by \cite{crowson2009technical}. Moreover, the hypothetical manipulation of cause-specific hazard rates in multi-state models has been also discussed by \cite{keiding2001multi} and most recently by \cite{gran2015causal}.

We found that, in contrast to the PAF as effect measure for baseline exposures, the proportion of attributable risk based on observable quantities cannot be described by the same statistical measure as the proportion of preventable cases if exposure was extinct. Based on the differentiation between explanatory, predictive and descriptive statistical modelling by \cite{shmueli2010explain}, we developed a clear concept of the two different estimands.
The former, $PAF_o(t)$, is a descriptive quantity that serves to understand the factual burden of exposure. The latter, $PAF_c(t)$, is an explanatory quantity and a causal extension of the PAF for time-dependent exposures. \cite{pearl2009causality} described the difference between causal and conventional modelling as the difference between seeing (descriptive point of view) and doing (causal point of view). In this sense, we categorized $PAF_o(t)$ as 'seeing' and $PAF_c(t)$ as 'doing'.

A prominent example which has led to a similar discussion as ours is the Cox PH model. By conditioning on survival up to a certain time point and assuming a time constant hazard ratio, the Cox PH model has a built-in selection bias similar to $PAF_o(t)$ (\cite{aalen2015does}). Nevertheless, it is one of the most relevant quantities that properly takes the temporal dynamics into account.

We used our formalization of $PAF_o(t)$ and $PAF_c(t)$ to explain misinterpreted results found in the literature (see \cite{bekaert2010adjusting}). Yet up to now, a distinction between causal effect measures and measures based on observable quantities is often only discussed in terms of confounding. However, our results suggest that in the time-dependent situation, confounding is not the crucial difference between the two modelling strategies. By properly distinguishing between estimands and estimators (\cite{akacha2017estimands}) and simplifying the estimation procedure for the PAF proposed by \cite{bekaert2010adjusting}, we showed that differences between adjusted and unadjusted estimates cannot be fully explained by confounding.
Instead, the differences were a consequence of the fact that the two estimation procedures resulted in two distinct statistical measures. Our conceptualization of the PAF clarifies which estimator is appropriate for which estimand. Given that they are correctly interpreted, the estimators proposed by \cite{bekaert2010adjusting} provide unbiased results and are an important contribution to the field. Especially the estimator $PAF_B^{IPW}(t)$ which we proofed to be an estimator of $PAF_c(t)$ allows for adjustment for time-varying confounding. Such adjustment is essential when estimating $PAF_c(t)$.

In this article, we approached the causally defined estimands of the PAF with conventional statistical modelling by using a multi-state model approach. Estimation of causal effects is only possible under strong and untestable assumptions. We mentioned the most important conditions for causal inference, namely exchangeability, consistency, and positivity. However, we did not go into detail and we did not formalize further conditions that might become necessary. These include, for example, correct model specification for parametric estimators and the assumption that the removal of exposure does not alter the distribution of other risk factors. 

Additionally to these untestable assumptions, we mentioned the Markov assumption which must be made to estimate transition probabilities of multi-state models with the Aalen-Johansen estimator. The Markov assumption facilitates estimation of transition probabilities that involve intermediate events. For example, to obtain a full picture of the course of infection and its consequences for infected patients, all transition probabilities of the extended illness-death model are of interest. 
In contrast, both estimands, $PAF_o(t)$ and $PAF_c(t)$, can be estimated without having to make the Markov assumption.
The PAF quantifies the burden of exposure on a population level and is not a measure for the consequences of exposure among already exposed patients. 

A causal approach to the estimation of population risks has been proposed by \cite{taubman2009intervening}. They use the g-formula to account for a time-dependent exposure and time-varying confounding. However, their exploration does not include data settings with competing risks.

In practice, time-dependent exposures are common but often avoided by e.g. summarizing data at the end of follow-up. Such data is then aggregatable in a fourfold table of exposure and outcome and the time-fixed definition of the PAF \eqref{eq:PAF_beni} can be used.
In the absence of censoring the estimand that results from this procedure is equal to $PAF_o(\tau)$, where $\tau$ denotes the end of follow-up (\cite{schumacher2007attributable, von2017basic}). Therefore, our conclusions on $PAF_o(t)$ apply equally to this frequently used approach.

More specifically, the estimand $PAF_o(\tau)$ has no causal interpretation even in the absence of confounding. The only exception is a data setting with time constant hazards, then we have that $PAF_o(\tau)=PAF_c(\tau)$ (\cite{von2017basic}). Nevertheless, $PAF_o(\tau)$ summarizes $PAF_o(t)$ at the end of follow-up and therefore serves as a description of the factually observed burden of exposure.

Other commonly used effect measures such as the RR, the RD, and the OR defined as measures of association are also based on $P(D(t)=1|E(t)=0)$. 
Hence, these associational effect measures also have no causal interpretation, even in the absence of confounding if the exposure occurs during follow-up.

Nevertheless, we provide a framework to interpret these estimands in a meaningful way. For example, despite having no causal interpretation $PAF_o(\tau)$ remains an estimand that validly quantifies the population-attributable burden of a harmful exposure.

Measures of association are said to be structurally biased if they differ from their causal analogues (\cite{greenland1999confounding, hernan2018causal}).
In the time-fixed data setting, the consistency and exchangeability assumptions are sufficient to be able to interpret measures of association as causal effect measures or vice versa to identify an estimand defined with counterfactual variables. (\cite{hernan2004definition}). However, in time-dependent data settings, previous exposure and outcome states influence the present states. The timing of exposure and outcome may not be a causal consequence of exposure. Nevertheless, time to exposure is a natural aspect of a time-dependent exposure.

The time-fixed PAF is said to be a causal quantity (\cite{mansournia2018population}). With regard to the different interpretation of the two estimands $PAF_c(t)$ and $PAF_o(t)$, we refrain from generalizing this statement to the more natural time-dependent data settings. 

For example, to obtain a profound understanding of the observable population-attributable burden and the proportion of preventable cases, both $PAF_o(t)$ and $PAF_c(t)$ should be considered. This has been demonstrated with a publicly available data example. We estimated the population-attributable burden of HAP and also the proportion of preventable ICU death cases if a hypothetical intervention prevented HAP for all patients during their complete ICU stays. 

Our results show that the timing of exposure acquisition contradicts causal interpretation of unconfounded associations and causal estimands may not capture natural aspects of exposure prevalence. 
Based on the PAF as effect measure, we argue that in time-dependent settings causal effect measures and measures of association must be regarded as distinct estimands with distinct interpretations. Moreover, disregarding the fundamental difference between time-fixed and time-dependent data settings easily leads to misinterpretations.

A proper distinction between effect measures based on observable quantities and those based on hypothetical ones helps to correctly interpret study results and to draw the correct conclusions. The estimands defined for time-dependent exposures based on observable quantities are often limited to a descriptive interpretation. In contrast, estimands based on hypothetical outcomes allow for a causal interpretation. However, an estimation of these quantities requires strong assumptions. Moreover, they cannot describe observable phenomena as the time-delay of exposure acquisition is not accounted for.

\newpage

\section*{Abbreviations}
\begin{tabular}{l l}  
$PAF_o$ & PAF as defined by \cite{benichou2001review} with observable variable \\ 
& for time-fixed data settings\\
& \\
$PAF_c$ & PAF as defined by \cite{sjolander2011estimation} with counterfactual variables\\
& for time-fixed data settings\\
& \\
$PAF_o(t)$ & PAF for time-dependent exposures and competing risks -\\ 
& defined with observable variables ( \cite{schumacher2007attributable})\\
& \\
$PAF_c(t)$ & PAF for time-dependent exposures and competing risks - \\
& defined with counterfactual variables \\
& \\
$\widehat{PAF}_B^{naive}(t)$ & Unadjusted estimator of the PAF as proposed by \cite{bekaert2010adjusting} \\ 
& \\
$\widehat{PAF}^{IPW}_B(t)$ &  Adjusted estimator of the PAF as proposed by \cite{bekaert2010adjusting}\\
 & \\
 & \\
 $P(D(t)=1|E(t)=0)$ & Specific version of the CPF\\
 &  (Death risk among unexposed as used in the definition of $PAF_o(t)$)\\
 & \\
 $P(D_0(t)=1)$ & Counterfactual death risk had all patients remained unexposed until time $t$ \\ 
 & (Death risk among unexposed as used in the definition of $PAF_c(t)$)\\
 & \\
$\widehat{F}_{\overline{0}1}^{naive}(t)$ & Unadjusted estimated death risk among unexposed as proposed by \cite{bekaert2010adjusting} \\
& \\
$\hat{F}^{IPW}_{\overline{0}1}(t)$ & Adjusted estimated death risk among unexposed as proposed by \cite{bekaert2010adjusting} \\
\end{tabular}

\section*{Availability of data and materials}
The R code of the data example is available in the supplementary material of this paper. The data is publicly available in the Comprehensive R Archive Network package (R) \textit{kmi} (\cite{kmi2017}).

\section*{Funding}
MvC was supported by the Innovative Medicines Initiative Joint Undertaking under grant agreement n [115737-2 – COMBACTE-MAGNET], resources of which are composed of financial contribution from the European Union’s Seventh Framework Programme (FP7/2007-2013) and EFPIA companies; MW has received funding from the German Research Foundation (Deutsche Forschungsgemeinschaft) under grant no. WO 1746/1-2.

\section*{Potential conflicts of interest}
Conflicts of interest for all authors: none.

\bibliography{databaseCOMBACTEMagnet}
\bibliographystyle{ieeetr}

\section*{Appendix}

\subsection{Definition of the counterfactual PAF as proposed by Bekaert et al.}

Bekaert et al. define an estimator of the PAF, which is based on estimated CIFs. 
Formally, they propose at each time $t$

\begin{equation}
\widehat{PAF}_B(t)=\frac{\hat{F}_1(t)-\hat{F}_{\overline{0}1}(t)}{\hat{F}_1(t)}.
\label{eqPAF_Bb}
\end{equation}

The quantity $\hat{F}_1(t)$ is the "estimated observed CIF of ICU mortality at time t" (\cite{bekaert2010adjusting}). 
The quantity $\hat{F}_{\overline{0}1}(t)$ is defined as the "estimated counterfactual CIF of ICU mortality at time $t$ under the no-infection path" (\cite{bekaert2010adjusting}). The no-infection path implies that the patient remained -- possibly contrary to fact -- infection-free during the complete follow-up time $(0,\tau]$. For a detailed explanation of the hypothetical infection paths we refer to \cite{bekaert2010adjusting}.

The estimand $F_{\overline{0}1}(t)$ is formally defined by \cite{bekaert2010adjusting} as
\begin{equation}
F_{\overline{0}1}(t)=P(T_{\overline{0}}\leq t, \epsilon_{\overline{0}}=1),
\end{equation}

where $T_{\overline{0}}$ is the counterfactual event time under the no-infection path, i.e. under the assumption that the patient would have remained  -- possibly contrary to fact -- infection-free, and $\epsilon_{\overline{0}}$ is the counterfactual outcome under the no-infection path. We have $\epsilon_{\overline{0}}=1$ if the patient had died in the ICU under the no-infection path and $\epsilon_{\overline{0}}=2$ if he had been discharged alive. Random censoring has not been considered.

The estimand $F_{\overline{0}1}(t)$ is equivalent to $P(D_0(t)=1)$. This is because $D_0(t)=1$ if the hypothetical event time is smaller or equal to the observation time $t$ ($T_{\overline{0}}\leq t$) \textit{and} the hypothetical type of the event is death in the ICU ($\epsilon_{\overline{0}}=1$).

Moreover, the estimand of $\hat{F}_1(t)$ can be formally defined as
\begin{equation}
F_1(t)=P(T\leq t, \epsilon=1),
\end{equation}
where $T$ is the observed event time of death or discharge and $\epsilon$ the observed type of outcome, with  $\epsilon=1$ if the patient died in the ICU and $\epsilon=2$ if having been discharged alive. The same notation has been also used by \cite{bekaert2010adjusting}. We have that $P(D(t)=1)=F_1(t)$.

\subsection{Proof of the equivalence of $\widehat{PAF}_o(t)$ and $\widehat{PAF}_B^{naive}(t)$}
In the following we show that $\widehat{F}_{\overline{0}1}^{naive}(t)$ is equivalent to the estimator of the death risk among unexposed, $\widehat{P}(D(t)=1|E(t)=0)$. Then, the equality of  $\widehat{PAF}_o(t)$ and $\widehat{PAF}_B^{naive}(t)$ is a direct consequence.

\subsubsection{Notations and definitions}
Like commonly done in causal inference, \cite{bekaert2010adjusting} properly differentiate between observable and counterfactual outcomes. As introduced in the previous section, the observable event time was defined by $T$ and the hypothetical event time, had the patient followed the no-infection path, by $T_{\overline{0}}$.

The no-infection path $\overline{0}=(0_1, ..., 0_\tau)$ is a potential infection history of a patient. The infection state of a patient is measured on a daily basis. At each day within $[1,\tau]$ the patients could potentially acquire an infection, given they are still in the ICU at that day. The no-infection path is the potential path where patients remain infection-free at every day during their ICU stay. A patient who has been discharged (dead or alive) without an HAI remains infection-free until the end of follow-up ($\tau$), as an HAI can be only acquired in the ICU.

The observable event indicator was denoted by $\epsilon$ (1 for death in the ICU, 2 for discharge alive) and the hypothetical one under the no-infection path by $\epsilon_{\overline{0}}$. The observable indicator of infection was $A_t$, which becomes one if the patient was observed to acquire an HAI within $(0,t]$.

The consistency assumption states that the hypothetical event time $T_{\overline{0}}$ and outcome $\epsilon_{\overline{0}}$ of patients who were observed to be unexposed until the end of their ICU stay is equal to their observed event time and state. Thus,
\begin{equation}
T_{\overline{0}}=T \quad \text{and} \quad \epsilon_{\overline{0}}=\epsilon, \quad \text{if } A(T)=0.\label{eq:cons}
\end{equation}

The random variables $T$, $A(t)$ and $\epsilon$ observed for the specific patient $i$ in the study cohort are denoted by $T_i$, $A_i(t)$ and $\epsilon_i$. The study cohort consists of $n$ patients.

The proof will be conducted by translating the estimator of Bekaert et al. into an estimator based on counting process notation.

\begin{figure}[hbtp!]
\centering
\includegraphics[width=13cm]{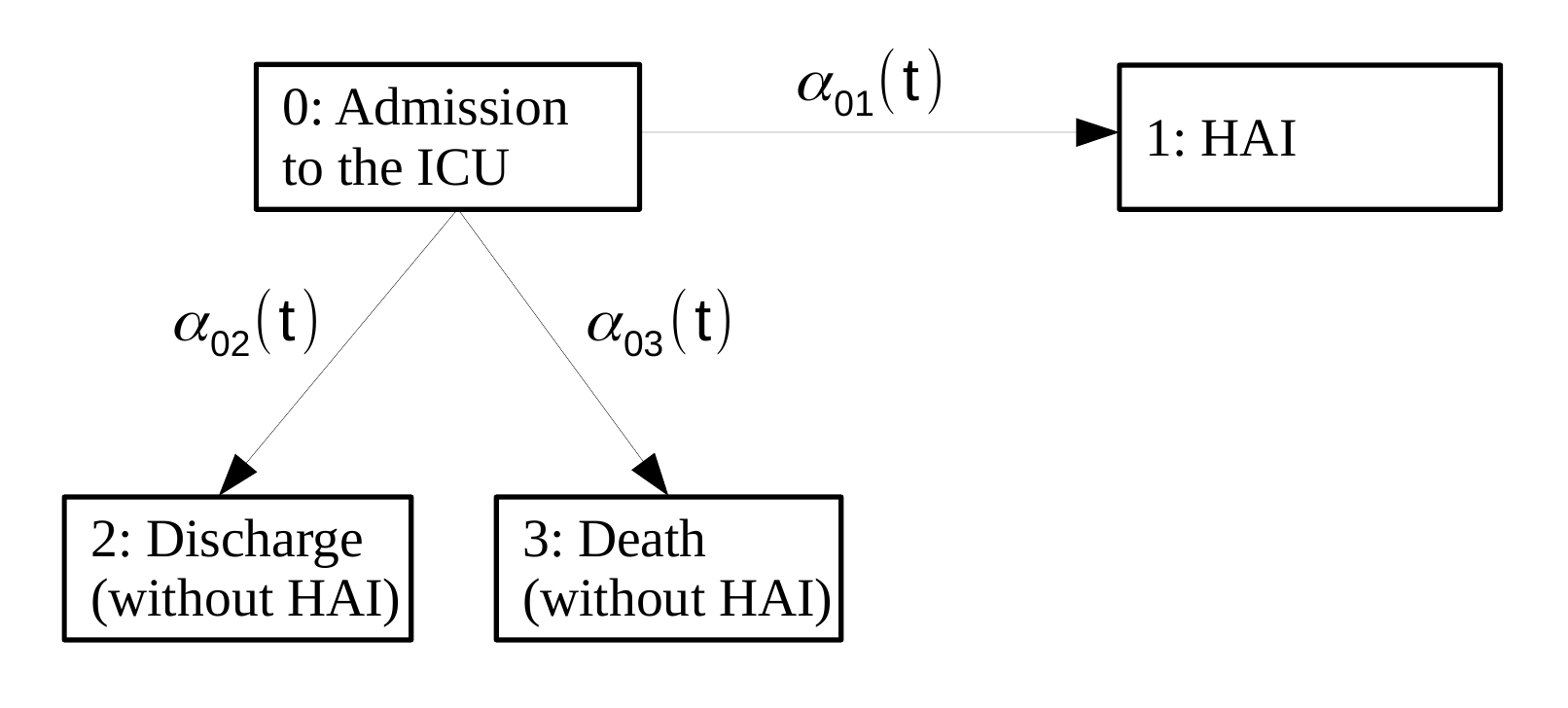}
 \caption{The competing risks model with hazard rates $\alpha_{01}(t)$, $\alpha_{02}(t)$, $\alpha_{03}(t)$.}
  \label{fig:CR_IllnessDeath}
\end{figure}

To simplify notation in the proof, we consider a competing risks process with three-states, as shown in Figure \ref{fig:CR_IllnessDeath}, instead of the extended illness-death model. By $X(t)$ we denote the state occupied at time $t$, $t\in [0,\tau]$, $X(t)\in \{0,1,2,3\}$. The defined counting process results from the extended illness-death model by disregarding States 4 and 5. This means that the patients entering State 1 remain in State 1. We denote the transition-probabilities of this model by $P_{kl_{CR}}(0,t)$.
Then, $P_{01_{CR}}(t)$ is simply the CIF of acquiring an infection. The transition probabilities $P_{02}(0,t)$ and $P_{03}(0,t)$ are the same in both models (i.e. $P_{02}(0,t)=P_{02_{CR}}(0,t)$ and $P_{03}(0,t)=P_{03_{CR}}(0,t)$).

The following proof could be equivalently conducted with the counting process defined by the extended illness-death model (Figure \ref{fig:extendedIllnessDeath}), since the CIF $P_{01_{CR}}(t)$ of the 3-state competing risks model (\ref{fig:CR_IllnessDeath}) is equal to the sum $P_{01}(t)+P_{04}(t)+P_{05}(t)$ of the extended illness-death model.
Thus, we have that 

\begin{align*}
P(D(t)=1|E(t)=0)&=\frac{P_{03}(0,t)}{1-(P_{01}(t)+P_{04}(0,t)+P_{05}(0,t))}\\
&=\frac{P_{03}(0,t)}{1-P_{01_{CR}}(0,t)}\\
&=\frac{P_{03_{CR}}(0,t)}{1-P_{01_{CR}}(0,t)}.
\end{align*}

Moreover, for each observation $i$, $i=1,...,n$, we introduce the following notation which is based on the book by \cite{beyersmann2011competing}):
\begin{align*}
N_{kli}(t)&:=\mathbbm{1}(i\text{ makes a transition from $k$ to $l$ in $[0,t]$})\\
Y_{ki}(t)&:=\mathbbm{1}(i \text{ is in $k$ at } t-),
\end{align*}
where $l,k \in \{0,1,2,3\}$, $t-$ denotes the time shortly before $t$. In the ICU data situations $t-$ can be interpreted as $t-1$, $t$ in days since ICU admission.

Furthermore, we define
\begin{align*}
N_{kl}(t)&:=\sum_{i=1}^n N_{kli}(t),\\
Y_{k}(t)&:=\sum_{i=1}^n Y_{ki}(t),
\end{align*}
where $N_{kl}(t)$ is the number of patients who made a transition from State $k$ to State $l$ in $[0,t]$ and $Y_k(t)$ is the number of patients in State $k$ shortly before $t$ (i.e. at $t-$).

As \cite{bekaert2010adjusting}, we assume complete follow-up. Then, the Aalen-Johansen estimators of the transition probabilities of the 3-state competing risks model are simple proportions (\cite{von2017basic})
\begin{equation*}
\hat{P}_{0l_{CR}}(t)=\frac{N_{0l}(t)}{n},
\end{equation*}
with $l\in\{1,2,3\}$.

\subsubsection{Proof of $\widehat{F}_{\overline{0}1}^{naive}(t)=\widehat{P}(D(t)=1|E(t)=0)$}
Now, we show mathematically that $\widehat{F}_{\overline{0}1}^{naive}(t)$ is equivalent to $\widehat{P}(D(t)=1|E(t)=0)$. Based on the 3-states competing risks model defined in 3.1, we have
\begin{equation}
\widehat{P}(D(t)=1|E(t)=0)=\frac{\hat{P}_{03_{CR}}(t)}{1-\hat{P}_{01_{CR}}(t)},
\end{equation}
where $\hat{P}_{03_{CR}}(t)$ is the CIF of death without HAI by time $t$ and $\hat{P}_{01_{CR}}(t)$ is the estimated probability to be infected by time $t$. Since we do not consider States 4 and 5, $\hat{P}_{01_{CR}}(t)$ is an estimator of the CIF of infection.

First, we rewrite $\widehat{F}_{\overline{0}1}^{naive}(t)$ using our counting process notation.
\cite{bekaert2010adjusting} defined
\begin{align*}
\widehat{F}_{\overline{0}1}^{naive}&=\frac{\sum_{s=1}^td_{\overline{0}1s}}{n_{\overline{0}t}}\\
&=\frac{\sum_{i=1}^n \mathbbm{1}(\epsilon_i(t)=1)\prod_{s=1}^t(\mathbbm{1}(A_i(s)=0)\mathbbm{1}(\epsilon_i(s-1)=0)+\mathbbm{1}(\epsilon_i(s-1)\neq0))}{\sum_{i=1}^n \prod_{s=1}^t(\mathbbm{1}(A_i(s)=0)\mathbbm{1}(\epsilon_i(s-1)=0)+\mathbbm{1}(\epsilon_i(s-1)\neq0))}\\
&:=\frac{\sum_{i=1}^n \mathbbm{1}(\epsilon_i(t)=1) \times p_i(t)}{\sum_{i=1}^n p_i(t)},
\end{align*}
with $p_i(t):=\prod_{s=1}^t(\mathbbm{1}(A_i(s)=0)\mathbbm{1}(\epsilon_i(s-1)=0)+\mathbbm{1}(\epsilon_i(s-1)\neq0))$.

Rewriting these quantities with our notation, we obtain
\begin{align*}
\mathbbm{1}(A_i(s)=0)&=1-N_{01i}(s)\\
\mathbbm{1}(\epsilon_i(s-1)=0)&=\mathbbm{1}(\epsilon_i(s-)= 0)=Y_{0i}(s)+Y_{1i}(s)\\
\mathbbm{1}(\epsilon_i(s-1)\neq 0)&=\mathbbm{1}(\epsilon_i(s-)\neq 0)=1-(Y_{0i}(s)+Y_{1i}(s)),\\
\mathbbm{1}(\epsilon_i(s)= 1)&=N_{03i}(s).
\end{align*}

Then,
\begin{align*}
p_i(t)&=\prod_{s\leq t} ((1-N_{01i}(s))(Y_{0i}(s)+Y_{1i}(s))+1-(Y_{0i}(s)+Y_{1i}(s)))\\
&=\prod_{s\leq t} (Y_{0i}(s)+Y_{1i}(s)-N_{01i}(s)Y_{0i}(s)-N_{01i}(s)Y_{1i}(s)+1-Y_{0i}(s)-Y_{1i}(s))\\
&=\prod_{s\leq t} (1-N_{01i}(s)Y_{0i}(s)-N_{01i}(s)Y_{1i}(s))\\
&=\begin{cases}
1,&\text{if $N_{01i}(t)=0$},\\
0, &\text{if $N_{01i}(t)\neq 0$}.
\end{cases}
\end{align*}
Since $N_{01i}(t)\neq 0$, it follows that there exists $s_0 \leq t$ ($s_0$ is the time of infection) such that $N_{01i}(s_0)=1$ and $Y_{0i}(s_0)=1$ and $Y_{1i}(s_0)=0$. Thus, $1-N_{01i}(s_0)Y_{0i}(s_0)-N_{01i}(s_0)Y_{1i}(s_0)=1-1-0=0$, implying that $p_i(t)=0$, since $s_0\leq t$.

Therefore, we can simply write
\begin{equation*}
p_i(t)=1-N_{01i}(t).
\end{equation*}
It follows that 
\begin{equation*}
n_{\overline{0}t}=\sum_{i=1}^n p_i(t)=\sum_{i=1}^n (1-N_{01i}(t))=n-N_{01}(t)
\end{equation*}
and
\begin{align*}
\sum_{s=1}^t d_{\overline{0}1s}&=\sum_{i=1}^n N_{03i}(t)(1-N_{01i}(t))\\
&=\sum_{i=1}^n N_{03i}(t)-N_{01i}(t)N_{03i}(t),\\
&=\sum_{i=1}^n N_{03i}(t)=N_{03}(t),
\end{align*}
since $N_{03i}(t)N_{01i}(t)=0$ for all $t$.

Summing up, we showed that
\begin{equation*}
\widehat{F}_{\overline{0}1}^{naive}(t)=\frac{\sum_{s=1}^t d_{\overline{0}1s}}{n_{\overline{0}t}}=\frac{N_ {03(t)}}{n-N_{01}(t)}=\frac{\hat{P}_{03_{CR}}(1)}{1-\hat{P}_{01_{CR}}(t)}=\widehat{P}(D(t)=1|E(t)=0).
\end{equation*}

\subsection{Proof of the equivalence of $\widehat{PAF}^{IPW}_B(t)=\widehat{PAF}_c(t)$}

Now we show that the IPW-adjusted estimator
\begin{equation*}
\hat{F}^{IPW}_{\overline{0}1}(t)=\frac{\sum_{i=1}^n \mathbbm{1}(\epsilon_i(t)= 1) W_{i\overline{0}t}}{\sum_{i=1}^nW_{i\overline{0}t}},
\end{equation*}
with
\begin{equation*}
W_{i\overline{0}t}=\prod_{s=1}^t\frac{(\mathbbm{1}(A_i(s)=0)\mathbbm{1}(\epsilon_i(s-1)=0)+\mathbbm{1}(\epsilon_i(s-1)\neq0)}{P(A_i(s)=0|(\epsilon_i(s-1), A_i(s-1))},
\end{equation*}
is an inverse-probability of censoring weighted (IPCW) CIF which results from the competing risks model (Figure \ref{fig:CR_IllnessDeath}) by treating infected patients as randomly censored. Since this estimator is equivalent to $\hat{P}_{03_0}(t)$ (see \ref{HorvitzCIF}), it follows that $\widehat{PAF}^{IPW}_B(t)=\widehat{PAF}_B(t)$.

\subsubsection{The Horvitz-Thompson like estimator of the CIF}\label{HorvitzCIF}

\cite{satten2001kaplan} showed the equivalence of the Kaplan-Meier estimator and a Horvitz Thompson like estimator of the survival probability in a classic survival model (see exemplary Figure \ref{fig:surv_IllnessDeath}).

Similarly, \cite{antolini2006crude} showed that the CIF of a competing risks model in the presence of random censoring can be estimated as a Horvitz-Thompson like estimator. This is an IPCW version of the CIF (\cite{antolini2006crude}).

We consider the competing risks model of Section \ref{sec:PAFBb} that results from the one in Figure \ref{fig:CR_IllnessDeath} by treating infected patients as randomly censored instead of accounting for infection as a competing risk to death and discharge without HAI (Figure \ref{fig:CRcens}).

Formally, we denote the counting process defined by this model by $X^{CR}$. Moreover, we define by $T_c$ a random censoring time and $T^{CR}:=min(T_*, T_c)$ with $T_*$ being the (potentially due to random censoring unobserved) event time of event type 2 or 3. Finally, $T^{CR}_i$ is the observed time $T^{CR}$ of patient $i$. (Thus, $N_{03i}(T_i^{CR})=\mathbbm{1}(T_i \leq t,\text{event of type 3 was observed for $i$})$. 

For the event death ($X^{CR}_{T^{CR}}=3$) with the solely competing event discharge alive and random censoring, the  Horvitz-Thompson like estimator of the CIF is 
\begin{equation*}
\hat{P}_{03}^{IPW}(t)=\frac{1}{n}\sum_{i=1}^{n}\frac{N_{03i}(T_i^{CR})}{\hat{S}_{T_c}(T_i^{CR}-)},
\end{equation*}
where $\hat{S}_{T_c}$ is the estimated censoring distribution which could be for example estimated by a Kaplan-Meier estimator. 

In the following, we use the equivalence $\hat{P}_{03_0}(t)=\hat{P}_{03}^{IPW}(t)$. However, the random censoring time $T_c$ is the (potentially unobserved) infection time denoted by $T_{01}$. This means that the censoring rate in Figure \ref{fig:CRcens} is equal to the infection rate in Figure \ref{fig:CR_IllnessDeath}.

\subsubsection{Proof of $\hat{F}^{IPW}_{\overline{0}1}(t)=\hat{P}_{03_0}(t)$}

To begin with we consider the 3-state competing risks model defined by Figure \ref{fig:CR_IllnessDeath}. We denote by $T_i$ the observed event time of patient $i$. Formally, $T_i=\inf(t:X_i(t)\in\{1,2,3\})$, where $X_i(t)$ is the state occupied by patient $i$ at time $t$ corresponding to the competing risks process in Figure \ref{fig:CR_IllnessDeath}. The hazard rates of the counting process have also been defined in Figure \ref{fig:CR_IllnessDeath}. Thus, $\alpha_{01}(t)$ is the cause-specific hazard rate of moving from State $0$ to State $1$ at time $t$. 

\begin{figure}[hbtp!]
\centering
\includegraphics[width=10cm]{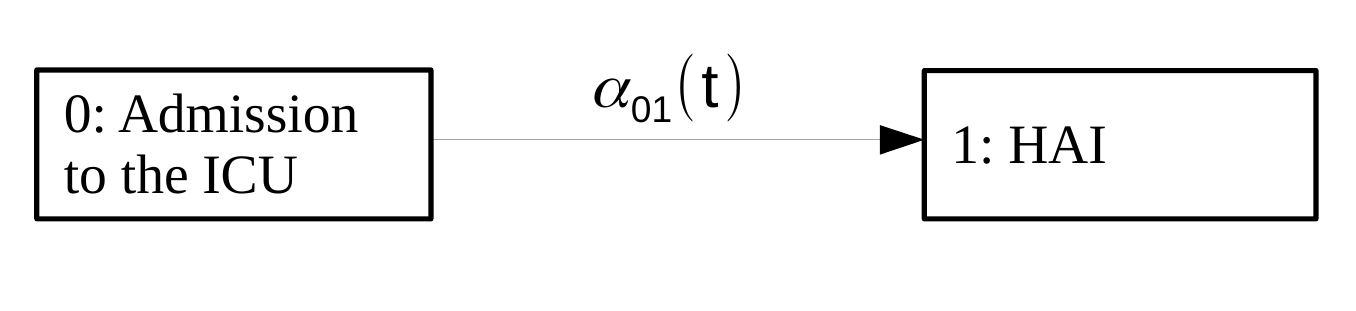}
 \caption{Classic survival model with hazard rate $\alpha_{01}(t)$.}
  \label{fig:surv_IllnessDeath}
\end{figure}

We have
\begin{align*}
W_{i\overline{0}t}&=\prod_{s \le t}\frac{p_i(t)}{P(A_i(s)=0|\epsilon_i(s-1), A_i(s-1))}\\
&\overset{(1)}{=}\prod_{s \le t}\frac{p_i(t)}{P(A_i(s)=0|\epsilon_i(s-1), A_i(s-1)=0)}\\
&\overset{(2)}{=}\prod_{s \le t\wedge T_i }\frac{p_i(t)}{P(A_i(s)=0|\epsilon_i(s-1)=0, A_i(s-1)=0)}\\
&=\prod_{s \le t\wedge T_i }\frac{p_i(t)}{P(\tilde{X}_i(s)\neq 1)|\tilde{X}_i(s-1)=0)}\\
&=\prod_{s \le t\wedge T_i }\frac{p_i(t)}{1-P(T\leq s, \tilde{X}_T=1|T>s-ds)}\\
&=\prod_{s \le t\wedge T_i }\frac{p_i(t)}{1-\alpha_{01}(s-ds)}\\
&\overset{(3)}{=}\frac{p_i(t)}{\prod_{s \le t\wedge T_i }(1-\alpha_{01}(s-ds)ds)}\\
&=\frac{p_i(t)}{\prod_{s \le (t\wedge T_i)-ds }(1-\alpha_{01}(s)ds)}\\
&\overset{(4)}{=}\frac{p_i(t)}{S_{01}((t\wedge T_i)-ds)}:=\frac{p_i(t)}{S_{01}((t\wedge T_i)-)},
\end{align*}
where $S_{01}(t):=P(T_{01}>t)$, with $T_{01}$ denoting the random survival time in a standard survival model with $\alpha_{01}(t)$ as the corresponding hazard rate. For an illustration of the model see Figure \ref{fig:surv_IllnessDeath}. Equation (1) follows since $p_i(t)=0$ if there exists $s_0\leq t$ such that $A_i(s_0-1)=1$. Equation (2) follows from $P(A_i(s)=0|(\epsilon_i(s-1)\neq 0, A_i(s-1)=0))=1$. Equation (3) is a result of $ds:=1$ (as $\alpha_{01}(s)$ is the daily rate of infection). Finally, equation (4) has been shown by e.g. Aalen et al. \cite{aalen2008survival} (Appendix A.1).

In other words,
\begin{equation*}
W_{i\overline{0}t}=\begin{cases}
0, \quad \text{if $N_{01i}(t)=1$},\\
\frac{1}{S_{01}(T_i-)},\quad  \text{if $N_{02i}(t)+N_{03i}(t)= 1$}\\
\frac{1}{S_{01}(t-)}, \quad \text{if $Y_{0i}(t)=1$}
\end{cases}
\end{equation*}

$S_{01}(t-)$ can be estimated with the Kaplan-Meier estimator.

Bekaert et al. use a pooled logistic regression model to estimate $\alpha_{01}(t)$. For example, if we use the most basic form of this model (no adjustment for time or and any covariates), we obtain the daily incidence rate of infection, i.e. 
\begin{equation*}
\frac{N_{01}(\tau)}{\text{Summed \# of patient days}}=\frac{N_{01}(\tau)}{\sum_{i=1}^n T_i}.
\end{equation*}

Furthermore, we have
\begin{align*}
\hat{F}_{\overline{0}1}(t)&=\frac{\sum_{i=1}^n \mathbbm{1}(\epsilon_i(t)= 1) W_{i\overline{0}t}}{\sum_{i=1}^nW_{i\overline{0}t}}\\
&=\frac{\sum_{i=1}^n N_{03i}(t) W_{i\overline{0}t}}{\sum_{i=1}^nW_{i\overline{0}t}}
\end{align*}
with
\begin{align*}
\sum_{i=1}^nW_{i\overline{0}t}&=\sum_{i=1}^n\frac{p_i(t)}{\hat{S}_{01}((t\wedge T_i)-)}\\
&=\sum_{i=1}^n\frac{N_{03i}(t)+N_{02i}(t)}{\hat{S}_{01}(T_i-)}+\sum_{i=1}^n\frac{Y_{0i}(t)}{\hat{S}_{01}(t-)}.\\
\end{align*}
Therefore, we can reformulate using the results of Section \ref{HorvitzCIF}
\begin{align*}
\hat{F}_{\overline{0}1}(t)&=\frac{\sum_{i=1}^n \frac{N_{03i}(t)}{\hat{S}_{01}(T_i-)}}{\sum_{i=1}^n\frac{N_{03i}(t)+N_{02i}(t)}{\hat{S}_{01}(T_i-)}+\sum_{i=1}^n\frac{Y_{0i}(t)}{\hat{S}_{01}(t-)}}\\
&=\frac{\frac{1}{n}\sum_{i=1}^n \frac{N_{03i}(t)}{\hat{S}_{01}(T_i-)}}{\frac{1}{n}\sum_{i=1}^n\frac{N_{03i}(t)+N_{02i}(t)}{\hat{S}_{01}(T_i-)}+\frac{1}{n}\sum_{i=1}^n\frac{Y_{0i}(t)}{\hat{S}_{01}(t-)}}\\
&= \frac{\hat{P}_{03_0}(t)}{\hat{P}(T^{CR}\leq t)+\hat{P}(T^{CR}>t)}\\
&=\frac{\hat{P}_{03_0}(t)}{1}=\hat{P}_{03_0}(t).
\end{align*}

To sum up, we showed that without consideration of confounding $\hat{F}_{\overline{0}1}(t)=\hat{P}_{03_0}(t)$. Thus, $\widehat{PAF}^{IPW}_B(t)=\widehat{PAF}_c(t)$.

\end{document}